\documentclass[12pt,letterpaper]{article}
\usepackage[left=1in,top=1in,right=1in,nohead]{geometry}
\usepackage{enumerate}
\usepackage{graphicx}
\usepackage{amsmath}
\usepackage{amsthm}
\usepackage{mathrsfs}
\usepackage{amssymb}
\usepackage{multirow}
\usepackage{color}
\usepackage{bm}
\usepackage{comment}
\usepackage{subfigure}
\usepackage{paralist}
\usepackage{jheppub}
\usepackage[utf8]{inputenc}

\newcommand{\grad}{\nabla}

\newcommand{\Lcal}{\mathcal{L}}

\newcommand{\eps}{\epsilon}

\newcommand{\T}{\mathcal{T}}

\newcommand{\fb}{\bar{f}}
\newcommand{\cb}{\bar{c}}
\newcommand{\nub}{\bar{\nu}}
\newcommand{\vb}{\bar{v}}
\newcommand{\Kb}{\overline{K}}
\renewcommand{\vec}[1]{\bm{#1}}

\renewcommand{\Im}{\mathrm{Im}}
\renewcommand{\Re}{\mathrm{Re}}

\def\bea#1\eea{\begin{align}#1\end{align}}
\def\be#1\ee{\begin{equation}#1\end{equation}}
\def\dd{\mathrm{d}}

\title{Towards Fluid Instabilities of Stationary Non-Killing Horizons}

\author[a]{Sebastian Fischetti}
\author[b]{and Benson Way}

\affiliation[a]{Theoretical Physics Group, Blackett Laboratory, Imperial College, London SW7 2AZ, UK}
\affiliation[b]{DAMTP, Centre for Mathematical Sciences, University of Cambridge, Wilberforce Road, Cambridge CB3 0WA, UK}

\emailAdd{s.fischetti@imperial.ac.uk}
\emailAdd{B.Way@damtp.cam.ac.uk}

\abstract{Flowing black holes are asymptotically locally AdS spacetimes that are stationary but have non-Killing horizons.  Holographically, they are dual to a steady-state heat flow in the boundary field theory.  We investigate the stability of these black holes in the limit in which they are well-described by the relativistic conformal Navier-Stokes equations.  More precisely, we study the quasi-normal modes of the linearized ideal fluid equations.  Though we find no unstable modes, there are an infinite number at finite transverse momentum which are arbitrarily long-lived.  This suggests the possibility that either non-modal effects or nonlinear interactions between these modes can give rise to new types of gravitational instabilities.}

\begin{document}

\maketitle
\flushbottom


\section{Introduction}
\label{sec:intro}

It is easy to conceive of steady-state systems that are out of global thermal equilibrium.  Consider, for example, the flow of heat between thermal reservoirs of different temperatures.  If the reservoirs have finite entropy, such a system will eventually equilibrate to global thermal equilibrium. However, if the reservoirs have infinite entropy, it is possible to maintain a steady-state heat flow between them.

In the context of the AdS/CFT correspondence~\cite{Mal97, Wit98a,GubKle98}, such systems are dual to an interesting class of asymptotically locally AdS black holes.  To see this, suppose a holographic field theory is put in such a steady-state configuration (with the infinite-entropy thermal reservoirs provided by \textit{e.g.}~nondynamical black holes or heat baths at asymptotic infinity).  One would conclude that the gravitational bulk dual must be stationary and contain a horizon of non-constant temperature, \textit{i.e.}~a stationary, non-Killing horizon.  We will refer to such geometries as ``flowing black holes''. Their existence was first motivated in~\cite{HubMar09}, and examples were later constructed numerically in~\cite{FisMar12,FigWis12}\footnote{See also~\cite{EmpMar13} for an analytic treatment that approximates a thin black string falling into a large black hole.}.

While these solutions could have been conceived without the aid of holography, there are a number of theorems in general relativity forbidding the existence of stationary black holes with non-Killing horizons~\cite{Hawking:1971vc,Hollands:2006rj,Moncrief:2008mr}.  These theorems, however, require the crucial assumption that the black hole horizon be compactly generated.  Indeed, we may understand the physical content of these theorems via standard black hole thermodynamics: heat flow generates entropy, and thus a flowing horizon can only be stationary if it contains asymptotic regions to which this entropy can be drained.  Thus any compact, stationary horizon must be in thermal equilibrium.

Regardless of any holographic interpretation, the novelty of flowing black holes makes them interesting objects in their own right.  However, they are difficult to construct, even numerically.  Fortunately, the fluid/gravity correspondence~\cite{BhaHub08,HubMin11} provides a limit in which flowing black holes are well-described by a hydrodynamic approximation. Specifically, when the (suitably-defined) local temperature scale of the horizon is much higher than any other characteristic inverse length scale, we expand the vacuum Einstein equation with negative cosmological constant in gradients.  Under such an expansion, the Einstein equation reduces to the relativistic Navier-Stokes equation describing a conformal fluid on the conformal AdS boundary.

This observation leads to a fascinating question: since stationary fluid flows are typically unstable at high enough Reynolds number~$Re$, can high-temperature flowing black holes exhibit analogous instabilities?  If so, high-temperature flowing black holes would exhibit phenomena akin to the end states of fluid instabilities, including other steady-state solutions, periodic solutions such as K\'arm\'an vortex streets, or fully-developed turbulence.  Such instabilities would be quite distinct from the currently known gravitational ones, such as Gregory-Laflamme~\cite{Gregory:1993vy} or superradiance~\cite{Zeldovich:1971,Zeldovich:1972,Starobinsky:1973scalar,
Starobinsky:1973,Press:1972zz}, though perhaps similar to the recently discovered instability of~\cite{GurJan16,JanJan16}.

In this Paper, we will begin to address the stability of flowing black holes in four dimensions by perturbing its dual three-dimensional ideal conformal fluid.  In particular, we will study the linear stability of a laminar flow over a gravitational potential well.  We will focus on computing quasi-normal modes (QNMs) of the linearized fluid equations, a procedure which in some nonrelativistic fluid contexts is called an Orr-Sommerfeld analysis.  While turbulent black hole horizons have been studied in the context of AdS/CFT~\cite{AdaChe12,CarLeh12,AdaChe13,GreCar13,CheLuc14}, all work to date has involved the evolution of initial data that is far from steady-state.  Instead, we are specifically interested in studying \textit{stability}, which inherently involves perturbing stationary solutions.

To be more concrete, our fluid lies on a three-dimensional background metric~$g_{ab}$.  In the hydrodynamic limit, the fluid is described by the local fluid temperature~$\T(x)$ and velocity field~$u^a(x)$.  The collective dynamics of these fields are governed by the conservation of the stress tensor,
\be
\label{eq:Tconservation}
\grad_a T^{ab} = 0,
\ee
where the stress tensor and dynamical variables are related through the constitutive relation for a conformal fluid,
\be
\label{eq:Tconstitutive}
T^{ab} = c_\mathrm{eff} \left[\T^3 \left(g^{ab} + 3 u^a u^b\right) + \Pi^{ab}\right].
\ee
Here,~$c_\mathrm{eff}$ is the effective CFT central charge, which is related to the bulk four-dimensional Newton's constant~$G_N$ and AdS length~$\ell$ by~$c_\mathrm{eff} = \ell^2/(16 \pi G_N)$.  The first term in~\eqref{eq:Tconstitutive} describes the stress tensor of an ideal conformal fluid; the second term~$\Pi^{ab}$ encodes viscous corrections which are organized in a gradient expansion in the fluid fields~$\T$,~$u^a$.  The flowing black holes in~\cite{FisMar12,FigWis12} were confirmed to be well-described by the effective fluid description~\eqref{eq:Tconservation} at sufficiently high temperatures.

\begin{figure}[t]
\centering
\subfigure[]{
\includegraphics[page=1]{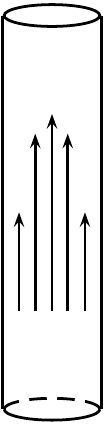}
\label{subfig:HagenPoiseuille}}
\hspace{2cm}
\subfigure[]{
\includegraphics[page=2]{Figures-pics}
\label{subfig:planePoiseuille}}
\hspace{2cm}
\subfigure[]{
\includegraphics[page=3]{Figures-pics}
\label{subfig:planeCouette}}
\caption{Three laminar, stationary solutions to the incompressible nonrelativistic Navier-Stokes equations.  \subref{subfig:HagenPoiseuille}: Hagen-Poiseuille flow, or flow through a round pipe.  \subref{subfig:planePoiseuille}: plane Poiseuille flow, or flow between two infinitely-extended plates.  \subref{subfig:planeCouette}: plane Couette flow, induced by placing a fluid between two infinitely-extended plates and dragging one relative to the other.  All three are unstable at high enough Reynolds number~$Re$, but only plane Poiseuille flow exhibits unstable modes in a linear stability analysis.}
\label{fig:fluidflows}
\end{figure}

In this Paper we study the stability of linear perturbations of the ideal fluid equations, \textit{i.e.}~\eqref{eq:Tconservation} with~$\Pi^{ab} = 0$.  However, we caution that while linear analysis accurately describes the stability of some solutions to the Navier-Stokes equation\footnote{Notable examples include Rayleigh-Benard convection of a plane horizontal layer of fluid heated from below, and Taylor-Couette flow between two rotating cylinders.}, there are many fluid instabilities that are not well-captured by this approach.  For example, consider three prototypical solutions of the nonrelativistic Navier-Stokes equations: Hagen-Poiseuille, plane Poiseuille, and plane Couette, shown schematically in Figure~\ref{fig:fluidflows}.  All three of these flows are unstable at high enough~$Re$, but in an Orr-Sommerfeld analysis only plane Poiseuille flow exhibits unstable modes~\cite{SchHen01}.  Moreover, experiments on plane Poiseuille flow found turbulence at a much lower~$Re$ than the one predicted by linear analysis~\cite{Orszag1971,Davies92,PatelHead,Carlson}.  Nevertheless, linear analysis has served as an important starting point for investigating the stability of these systems.

Like in Hagen-Poiseuille and Couette flow, we will find no unstable modes in our conformal fluid.  However, the modes we do find can have arbitrarily small damping at finite transverse momentum, which may possibly lead to instabilities through nonlinear coupling or non-modal effects~\cite{TreTre93}.  The former of these could potentially be tied to the parametric resonance instability of near-extremal Kerr~\cite{YanZim14}, wherein nonlinear coupling between long-lived modes allows energy to transfer between them.

This Paper is structured as follows.  In Section~\ref{sec:QNM}, we review the derivation of QNMs as poles of the retarded Green's function.  In Section~\ref{sec:idealfluid}, we compute the QNMs of the linearized ideal conformal fluid equations on a stationary background flow over a gravitational potential.  We will consider both an analytically tractable toy model where the gravitational potential is taken to be a finite step well, and also a smooth well which we treat numerically.  In Section~\ref{sec:viscousfluid}, we make some comments on the effect of including viscous corrections, and we discuss our results in Section~\ref{sec:discussion}.


\section{Review of Quasinormal Modes and Linear Response}
\label{sec:QNM}

Let us review the construction of quasi-normal modes (QNMs), with particular emphasis on deriving the boundary conditions that they must obey\footnote{We are indebted to Harvey Reall and Jorge Santos for illuminating discussions on some of the issues presented in this Section, specifically regarding the analytic structure of Green's functions.}.  In many gravitational contexts, solutions to linearized equations asymptotically behave (at large spatial coordinate~$x$, say) as
\be
\label{eq:outgoing}
e^{-i\omega t}\left(A^+ e^{i\omega x} + A^- e^{-i\omega x} \right) = A^+ e^{-i\omega(t - x)} + A^- e^{-i\omega(t + x)}
\ee
for complex frequency $\omega$.  These solutions are left-moving and right-moving waves traveling at the speed of light.  To construct QNMs, one usually imposes asymptotic outgoing boundary conditions by setting~$A^- = 0$.  These conditions are consistent with the intuition that QNMs physically govern excitations that eventually leave a globally hyperbolic patch.  However, as can be seen from~\eqref{eq:outgoing}, outgoing modes with~$\Im(\omega) < 0$ diverge at spatial infinity, so decaying QNMs should not be individually thought of as physical excitations.  Rather, their physical purpose is to bound the intermediate-time evolution of initial data of compact support (see \textit{e.g.}~\cite{KokSch99} for details).

In our analysis, we will instead encounter situations where the asymptotic solutions behave as
\be
A^+ e^{-i\omega t + i K^+(\omega) x} + A^- e^{-i\omega t + i K^-(\omega) x}
\ee
for some complex wave numbers~$K^\pm(\omega)$ which are nonlinear in~$\omega$.  In this case, it is no longer clear what is meant by ``outgoing''.  Therefore, rather than appealing to such a notion, we will instead study quasi-normal modes through their definition as the poles of retarded Green’s functions.

\subsection{Linear response in second-order systems}

Let us review the theory of linear response by following~\cite{KokSch99,BerCar09} for familiar examples.  Consider the linear, hyperbolic second-order PDE
\be
\label{eq:hyperbolicschematic}
\left[\frac{\partial^2}{\partial t^2} - \frac{\partial^2}{\partial x^2} + V(x)\right]\Phi = 0
\ee
for some potential~$V(x)$ and spatial coordinate~$x$ with infinite extent.  Here,~$\Phi$ may be any field, though for pedagogical purposes we will later take~$\Phi$ to be a Klein-Gordon field.

The purpose of a linear analysis is to understand how compactly-supported initial data evolves under~\eqref{eq:hyperbolicschematic}.  It is thus convenient to introduce the Laplace transform of~$\Phi$,
\be
\phi_\omega(x) \equiv \int_0^\infty \Phi(t,x) e^{i\omega t} \, \mathrm{d}t,
\ee
where instead of the usual Laplace transform variable~$s$ we have used~$s = -i\omega$ for later convenience.  As long as~$\Phi(t,x)$ is bounded in time (and even if it diverges more slowly than exponentially), the above transform converges in the upper half-plane~$\Im(\omega) > 0$.  Then in terms of~$\phi_\omega(x)$,~\eqref{eq:hyperbolicschematic} becomes
\be
\label{eq:inhomowave}
\Lcal \phi_\omega \equiv \left[\frac{d^2}{dx^2} + \Big(\omega^2 - V(x)\Big)\right]\phi_\omega = i\omega \Phi(0,x) -\dot{\Phi}(0,x) \equiv \mathcal{J}_\omega(x),
\ee
where a dot denotes differentiation with respect to~$t$.  Specifying initial data~$(\Phi(0,x),\dot{\Phi}(0,x))$ gives the source term~$\mathcal{J}_\omega(x)$, after which~\eqref{eq:inhomowave} can be solved with suitable boundary conditions to yield~$\phi_\omega$.  The resulting solution can be written in terms of the retarded Green's function~$G_\omega(x;x')$ of~$\Lcal$:
\be
\phi_\omega(x) = \int_{-\infty}^\infty G_\omega(x;x') \mathcal{J}_\omega(x') \, \mathrm{d}x'.
\ee
Finally, the time-domain solution~$\Phi(t,x)$ is obtained by inverting the Laplace transform,
\be
\label{eq:inverselaplace}
\Phi(t,x) = \frac{1}{2\pi}\int_{-\infty + ic}^{\infty + ic} \phi_\omega(x) e^{-i\omega t} \, \mathrm{d}\omega,
\ee
where~$c > 0$ is chosen to be large enough so that the contour of integration lies above any singularities of~$\phi_\omega$ in the complex~$\omega$-plane.

To identify general features of the time evolution of~$\Phi$ in response to some initial data, it is useful to consider deforming the contour of integration in the Laplace transform~\eqref{eq:inverselaplace} into the lower half-plane.  Generically,~$\phi_\omega(x)$ will not be analytic in the entire complex~$\omega$-plane, and the time evolution of~$\Phi(t,x)$ can be interpreted in terms of these non-analyticities.  In fact, it is clear that the analytic structure of~$\phi_\omega(x)$ in the complex~$\omega$-plane is inherited from the analytic structure of~$G_\omega(x;x')$, and thus to understand these non-analyticities we only need to understand those of the retarded Green's function.  

To that end, recall that given two linearly independent solutions~$\phi^\pm_\omega$ to the homogeneous equation~$\Lcal\phi_\omega^\pm = 0$, a Green's function of~$\Lcal$ can be constructed as
\be
\label{eq:Greensfunction}
G_\omega(x;x') = \frac{1}{W[\phi_\omega^-,\phi_\omega^+](x')} \left[\Theta(x-x') \phi_\omega^-(x') \phi_\omega^+(x) + \Theta(x'-x) \phi_\omega^-(x) \phi_\omega^+(x')\right],
\ee
where~$W[\phi_\omega^-,\phi_\omega^+](x')$ is the Wronskian\footnote{The particular differential operator~$\Lcal$ defined in~\eqref{eq:inhomowave} contains no first-derivative term, and thus Abel's identity implies that the Wronskian~$W[\phi_\omega^-,\phi_\omega^+](x')$ is in fact constant in~$x'$.  For a more general differential operator, it need not be.} of~$\phi_\omega^-$ and~$\phi_\omega^+$ and~$\Theta(x)$ is the Heaviside theta function.  Because~$G_\omega$ is a retarded Green's function, it must be bounded in space so that evolution of compactly supported initial data remains bounded.  Thus the linearly independent functions~$\phi_\omega^\pm$ must be chosen to vanish at~$x \to \pm \infty$ whenever~$\Im(\omega) > 0$.  The Green's function so obtained can then be analytically continued to the lower half-plane, and its analytic structure will determine the time evolution of~$\Phi(t,x)$ via the inverse Laplace transform~\eqref{eq:inverselaplace}.  As we will see, it is the requirement that~$\phi_\omega^\pm$ vanish at~$x \to \pm \infty$ for~$\Im(\omega) > 0$ that determines the boundary conditions defining QNMs.

\newpage
\noindent\underline{\textit{A massless scalar field}}\\

\noindent To make this construction more concrete, consider the simple case where~$\Phi$ is a massless scalar field on an asymptotically flat spacetime (\textit{e.g.}~on a Schwarzschild spacetime, in which case we take~$x$ to be the Regge-Wheeler tortoise coordinate~$r_*$).  Then~$V(x)$ vanishes asymptotically, and the asymptotic solutions to the homogeneous equation~$\Lcal\phi_\omega = 0$ behave as
\be
\phi_\omega(x) \sim A^+ e^{i \omega x} + A^- e^{-i \omega x}.
\ee
Requiring~$G_\omega(x;x')$ to decay asymptotically for~$\Im(\omega) > 0$ thus requires that we choose~$\phi_\omega^\pm(x \to \pm \infty) \sim e^{\pm i \omega x}$.  This uniquely specifies~$\phi_\omega^\pm$ everywhere in~$x$, and allows the construction of~$G_\omega(x;x')$.

\begin{figure}[t]
\centering
\subfigure[]{
\includegraphics[page=4,width=0.3\textwidth]{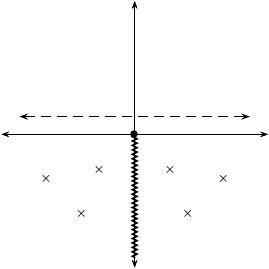}
\label{subfig:Ganalytic}
}
\hspace{1cm}
\subfigure[]{
\includegraphics[page=5,width=0.3\textwidth]{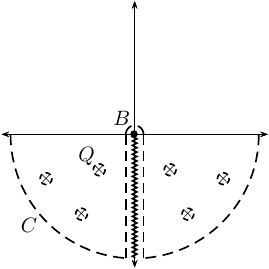}
\label{subfig:deformedcontour}
}
\caption{The analytic structure of the massless scalar field retarded Green's function~$G_\omega$~\eqref{eq:Greensfunction} in the complex~$\omega$-plane.  In the upper half plane, $G_\omega$ is analytic, but in the lower half plane, it contains poles (denoted by~$\times$) and generically a branch point at~$\omega = 0$.  Here we take the associated branch cut to lie along the negative imaginary axis.  \subref{subfig:Ganalytic}: the contour of integration for the inverse Laplace transform~\eqref{eq:inverselaplace} is shown as a dashed line.  \subref{subfig:deformedcontour}: by deforming the contour into the lower half-plane, the inverse Laplace transform can be divided into three contributions.  These contributions consist of a semicircle~$C$ at infinity which gives the (short time) prompt response, a discrete sum~$Q$ over the poles of~$G_\omega$ which give the (intermediate-time) quasinormal ringing, and the branch cut~$B$ which gives the (late-time) power-law tail.}
\label{fig:Laplace}
\end{figure}

The resulting analytic structure of~$G_\omega(x;x')$ was studied by Leaver~\cite{Lea86} (see~\cite{ChiLeu95} for a more general treatment). Typically,~$G_\omega(x;x')$ has branch points at~$\omega = 0$ and~$\omega = \infty$, and poles that lie in either of the lower complex quadrants, as shown schematically in Figure~\ref{fig:Laplace}.  When the contour of integration for the inverse Laplace transform (shown in Figure~\ref{subfig:Ganalytic}) is deformed into the lower half-plane, it can be split into three pieces (shown in Figure~\ref{subfig:deformedcontour}):
\begin{itemize}
	\item A semicircle~$C$ at infinity, encoding the \textit{prompt response} of the system.  This is the short-time response of the system to the initial compactly-supported perturbation as it propagates along the light cone, and is in fact the only contribution to the Green's function in flat space.
	\item The poles~$Q$ in the lower quadrants, corresponding to the \textit{quasi-normal modes} of the system.  These govern the system at intermediate times, when it exhibits exponentially decaying ``ringing''.
	\item The branch cut contribution~$B$ (when a branch point exists), giving the \textit{late-time power-law tails} that follow the quasi-normal ringing.  These power-law tails mostly result from the portion of the integration contour lying near the branch point at~$\omega = 0$.
\end{itemize}

\begin{figure}[t]
\centering
\includegraphics[page=6,width=0.3\textwidth]{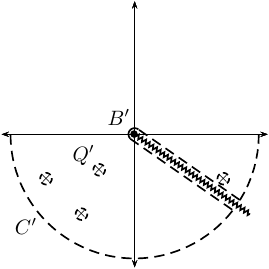}
\caption{The location of the branch cut corresponding to the branch point at~$\omega = 0$ of~$G_\omega$ need not lie along the negative imaginary axis. Here, we show a possible alternative choice of cut, as well as the associated integration contour for~\eqref{eq:inverselaplace}.  Note that with this choice of branch cut, poles that appeared in Figure~\ref{fig:Laplace} now lie on a secondary Riemann surface (behind the cut), and therefore do not contribute to the discrete sum~$Q'$.  Instead, they are contained in the branch cut contribution~$B'$, so that the resulting time evolution~$\Phi(t,x)$ is unchanged.}
\label{fig:differentcut}
\end{figure}

The perturbation~$\Phi(t,x)$ can thus be decomposed intro three pieces corresponding to each of these contributions:
\be
\Phi(t,x) = \Phi^C(t,x) + \Phi^Q(t,x) + \Phi^B(t,x).
\ee
While the locations of poles and branch points of~$G_\omega$ are fixed, note that there is some arbitrariness in where to draw the branch cut connecting the branch points\footnote{In fact, it may be possible to argue that there is a ``natural'' choice of branch cut associated with the fact that if the domain is compactified, the cut should resolve into a line of closely-spaced poles.  However, within the context of a single infinite-domain problem there is nothing to specify where the cut should go.}.  While it is conventional to take the branch cut to lie along the negative imaginary axis, it is permissible to choose it to lie anywhere else in the lower half-plane,~\textit{e.g.}~as in Figure~\ref{fig:differentcut}.  It is even possible to choose the cut so that some poles of~$G_\omega$ lie on a secondary Riemann surface, which would move the contribution of these poles from~$\Phi^Q(t,x)$  to~$\Phi^B(t,x)$.  However, as long as the integration contour for~\eqref{eq:inverselaplace} never crosses branch points or poles, the final expression for~$\Phi(t,x)$ will be unchanged. Thus the freedom to change the location of the branch cut corresponds to the ambiguity in defining~$\Phi^Q$ and~$\Phi^B$, which can be interpreted as an ambiguity in the notion of ``intermediate time''.

The QNMs are appealing to study because they can be extracted from the linear operator~$\mathcal{L}$ relatively easily.  From~\eqref{eq:Greensfunction}, the QNMs correspond to those~$\omega$ for which~$\phi_\omega^-$ and~$\phi_\omega^+$ become linearly dependent (\textit{i.e.}~their Wronskian vanishes).  For the massless scaler, these modes therefore correspond to solutions of~$\mathcal{L} \phi_\omega^\mathrm{QNM} = 0$ that behave at infinity as
\be
\label{eq:falloff}
\phi^\mathrm{QNM}_\omega(x \to \pm \infty) \sim e^{\pm i \omega x}.
\ee
Note that these are precisely the outgoing boundary conditions~\eqref{eq:outgoing} stated above.  However, we emphasise that the purpose of this discussion was to highlight that from the perspective of linear response and Green's functions, there is no consideration of the meaning of ``outgoing".  Instead, these boundary conditions are imposed by the requirement that the Green's functions exhibit the appropriate spatial decay when $\Im(\omega)>0$.
\\\\
\noindent\underline{\textit{A massive scalar field}}\\

\begin{figure}[t]
\centering
\includegraphics[page=7,width=0.3\textwidth]{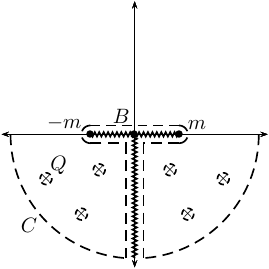}
\caption{The analytic structure of the retarded Green's function~$G_\omega$ of the massive scalar field in the complex~$\omega$-plane.  In addition to the usual poles and branch point at~$\omega = 0$, there are two new branch points at~$\omega = \pm m$.  We connect them with a cut along the real axis, though in principle the cut can be taken to lie anywhere in the lower half-plane.  The contour of integration for the inverse Laplace transform~\eqref{eq:inverselaplace} is again drawn as a dashed line; note that the contribution~$B$ from the branch cut now contains an extended portion along the real axis.}
\label{fig:massivecuts}
\end{figure}

\noindent We will find that our fluid system exhibits a more complicated analytic structure in the complex $\omega$ plane than the massless scalar field.  To illustrate this with a familiar example, take~$\Phi$ to be a massive scalar field with mass~$m$ on an asymptotically flat spacetime, as studied in \textit{e.g.}~\cite{KonZhi04,Zhi06,KonZhi06}.  Then asymptotically,~$V(x) \to m^2 \neq 0$, and thus the asymptotic solutions to the homogeneous equation~$\Lcal \phi_\omega = 0$ behave like
\be
\phi_\omega(x) \sim A^+ e^{i \sqrt{\omega^2 - m^2} \, x} + \sim A^- e^{- i \sqrt{\omega^2 - m^2} \, x}.
\ee
Unlike the massless case, these solutions introduce branch points at~$\omega = \pm m$, which prevents them from being cleanly interpreted as ``outgoing'' or ``incoming''.  Analyticity in the upper half-plane requires us to take the associated branch cut either on or below the real axis; we make the choice shown in Figure~\ref{fig:massivecuts}. Spatial decay at infinity whenever~$\Im(\omega) > 0$ then fixes the boundary conditions to impose on the solutions~$\phi_\omega^\pm$ from which we construct the Green's function:
\be
\label{eq:phipmmassive}
\phi_\omega^\pm(x \to \pm \infty) \sim e^{\pm i \omega \sqrt{1-m^2/\omega^2} \, x},
\ee
where we take the branch cut of the square root to be the standard one (\textit{i.e.}~the branch cut lies where~$1-m^2/\omega^2 < 0$).  Then the inverse Laplace transform~\eqref{eq:inverselaplace} can again be decomposed into three contributions: one from the semicircle~$C$, one from the set of QNMs~$Q$, and one from the~T-shaped branch cut~$B$, as illustrated in Figure~\ref{fig:massivecuts}.  As in the massless case, there is substantial freedom in the choice of where to locate the branch cuts.  For instance, they could also be chosen to extend into the lower half-plane.  In some sense, our choice is natural in that it is a small deviation from the branch cut structure in the massless case above.  We reiterate that despite this freedom, any choice gives the same time evolution~$\Phi(t,x)$.

It is now clear that the poles of~$G_\omega(x;x')$ correspond to those~$\omega$ at which there exists a solution to~$\Lcal \phi^\mathrm{QNM}_\omega = 0$ that behaves as
\be
\phi^\mathrm{QNM}_\omega(x \to \pm \infty) \sim e^{\pm i \omega \sqrt{1-m^2/\omega^2} \, x}.
\ee
Though these boundary conditions cannot be interpreted cleanly as ``outgoing'', they follow straightforwardly from an analytic continuation of the retarded Green's function.

\subsection{Generalization to higher differential order}

We will find that our fluid system yields a third-order differential equation, so we conclude this Section by extending the above analysis to higher differential order.  The essence of the discussion is unchanged, so we proceed schematically.  Consider the~$n^\mathrm{th}$-order linear differential operator
\be
\Lcal = \frac{d^n}{dx^n} + \sum_{i=0}^{n-1} q_i(x) \frac{d^i}{dx^i},
\ee
and take~$x$ to have infinite extent and assume that the~$q_i(x)$ become constant at large~$x$.  Then asymptotically, the~$n$ linearly independent solutions to the homogeneous problem~$\Lcal\phi = 0$ behave as~$e^{\lambda_i x}$ for some exponents~$\lambda_i$ (which will depend on the asymptotic values of the~$q_i$).  Generically, the~$\lambda_i$ will be distinct and have non-vanishing real part, and each of the~$e^{\lambda_i x}$ vanishes at either~$x \to -\infty$ or~$x \to \infty$.  Let us therefore split up these linearly independent solutions by the boundary at which they vanish:
\begin{subequations}
\label{eqs:phibndryconditions}
\bea
\phi_i^+(x \to \infty) &\sim e^{\lambda_i x} \to 0 \mbox{ for } i = 1, \ldots, p, \\
\phi_i^-(x \to -\infty) &\sim e^{\lambda_i x} \to 0 \mbox{ for } i = p+1, \ldots, n,
\eea
\end{subequations}
where we use the superscripts~$\pm$ when useful as reminders of where the~$\phi_x^\pm$ vanish.  Note that these conditions are not sufficient to fix the~$\phi_i^\pm$ uniquely, but they are sufficient to ensure that (as long as the~$\lambda_i$ are all distinct) the sets~$\{\phi^+_i\}$ and~$\{\phi^-_i\}$ are each linearly independent\footnote{In the degenerate case where two of the~$\lambda_i$, say~$\lambda_1$ and~$\lambda_2$, coincide, we simply take~$e^{\lambda_1 x}$ and~$x e^{\lambda_1 x}$ to be linearly independent asymptotic solutions.  This approach easily generalizes to when more than two~$\lambda_i$ coincide.}.

It is then straightforward to construct the generalization of the Green's function~\eqref{eq:Greensfunction} via \textit{e.g.}~the method of undetermined coefficients.  We obtain
\begin{multline}
\label{eq:generalGreens}
G(x;x') = \frac{(-1)^n}{W(x')}\left[\Theta(x-x') \sum_{i=1}^p (-1)^i W_i(x') \phi^+_i(x) \right. \\ \left. - \Theta(x'-x) \sum_{i = p+1}^n (-1)^i W_i(x') \phi^-_i(x)\right],
\end{multline}
where~$W(x')$ is the Wronskian of all the~$\{\phi_i\}$, while~$W_j(x')$ is the Wronskian of the~$\{\phi_i\}~\setminus~\phi_j$:
\begin{subequations}
\bea
W(x') &= \left| \begin{matrix} \phi_1 & \phi_2 & \cdots & \phi_n \\
							\phi'_1 & \phi'_2 & \cdots & \phi'_n \\
							\vdots & \vdots & \ddots & \vdots \\
							\phi_1^{(n-1)} & \phi_2^{(n-1)} & \cdots & \phi_n^{(n-1)} \end{matrix} \right|,
\\
W_i(x') &= \left| \begin{matrix} \phi_1 & \cdots & \phi_{i-1} & \phi_{i+1} & \cdots & \phi_n \\
							\vdots & \ddots & \vdots & \vdots & \ddots & \vdots \\
							\phi_1^{(n-2)} & \cdots & \phi_{i-1}^{(n-2)} & \phi_{i+1}^{(n-2)} & \cdots & \phi_n^{(n-2)}
\end{matrix} \right|.
\eea
\end{subequations}
This Green's function decays at infinity and is invariant under redefinitions of the~$\phi_i$ that leave the conditions~\eqref{eqs:phibndryconditions} invariant.  It is thus the correct Green's function for constructing solutions to the inhomogeneous differential equation~$\Lcal\phi = \mathcal{J}$ with~$\phi$ subject to vanishing boundary conditions at infinity.  As a check, note that it reproduces~\eqref{eq:Greensfunction} when~$n = 2$,~$p = 1$.

Now, if the differential operator~$\Lcal$ arose from the Laplace transform of some hyperbolic PDE (as in~\eqref{eq:inhomowave}), then the linearly independent solutions~$\phi_i$ -- and therefore also~$G(x;x')$ -- will depend on the Laplace variable~$\omega$.  To obtain QNMs, we then proceed just as in the second-order case.  First, in the upper half-plane~$\Im(\omega) > 0$, the linearly independent solutions~$\{\phi_i\}$ are defined by the asymptotic decay conditions~\eqref{eqs:phibndryconditions}.  These in turn are used to construct the Green's function~$G(x;x')$, which will be analytic everywhere in the upper half-plane~$\Im(\omega) > 0$.  Then~$G(x;x')$ is analytically continued to the lower half-plane, and its poles are identified as the QNMs.  From~\eqref{eq:generalGreens}, these poles correspond to zeros of the Wronskian~$W(x')$, or those~$\omega$ for which the~$\{\phi_i\}$ are not linearly independent.  But by construction, each of the sets~$\{\phi_i^+\}$ and~$\{\phi_i^-\}$ are linearly independent (assuming all the~$\lambda_i$ are distinct), and thus~$W(x')$ can only vanish when there exists a solution which behaves like the~$\phi_i^-$ as~$x \to -\infty$ and like the~$\phi_i^+$ as~$x \to \infty$.  These are the boundary conditions that define the QNMs.


\section{Linear Perturbations of the Ideal Fluid}
\label{sec:idealfluid}

In this section, we study the QNMs of perturbations of an ideal, laminar flow of a conformal fluid.  We describe the background flow, then consider perturbations first for a gravitational finite step well, then for a smooth gravitational well.

\subsection{Stationary Background Flows}
\label{subsec:stationary}

We consider fluid flows in the background spacetime
\be
\label{eq:metric}
\dd s^2 = -f(x)^2\dd t^2+\dd x^2+\dd y^2,
\ee
where $f$ is some function with $f(\pm\infty)=1$.  Recall that when~$|f(x)-1| \ll 1$, we can think of~$\Phi_N(x) \equiv f(x) -1$ as a Newtonian potential.  We will therefore often refer to~$f(x)$ as the gravitational potential through which the fluid flows.

We take the background fluid flow~$u_0^\mu$, normalized to~$u_0^2 = -1$, to be a stationary laminar flow in the~$x$-direction with translational invariance in~$y$:
\be
\label{eq:idealansatz}
u_0^\mu = \frac{1}{\sqrt{f(x)^2-v_0(x)^2}} \left(1, v_0(x), 0\right), \qquad \mathcal \T_0 = \T_0(x),
\ee
which has the feature that~$v_0(x) = u_0^x/u_0^t$ is a conformal invariant\footnote{Note that~$\T$ is related to the local fluid temperature~$T$ by~$\T \equiv 4\pi T/3$.}.  Without loss of generality, we take the flow to be in the positive-$x$ direction. Let us therefore define~$v_\infty \equiv v_0(-\infty)$ and~$\T_\infty \equiv \T_0(-\infty)$ to be the upstream fluid velocity and temperature, respectively.  Any nontrivial potential~$f(x)$ will also come equipped with a characteristic length scale~$L$ over which it varies, so the dimensionless parameters characterizing the flow are~$v_\infty$ and~$L \T_\infty$.  But recall from the discussion in Section~\ref{sec:intro} that the latter of these controls the gradient expansion, with the hydrodynamic approximation requiring~$L\T \gg 1$.  Then since the ideal fluid equations come from the leading-order terms in the gradient expansion, the ideal fluid flow depends on~$L \T_\infty$ only as an overall normalization for~$L\T_0(x)$.  Thus the only nontrivial parameter that characterizes such flows is~$v_\infty$.

For a background flow that varies only in~$x$, one of the fluid equations can be immediately integrated at any order in the hydrodynamic expansion,
\be
\grad_\mu T^{\mu t} = 0 \Rightarrow T^{xt} = \frac{3\T_\infty^3 v_\infty}{f(x)^3(1-v_\infty^2)},
\ee
where the constant of proportionality was fixed by recalling that~$f(x\to -\infty) = 1$ and that the viscous terms~$\Pi^{ab}$ vanish asymptotically.

For the ideal fluid, the remaining equations can be solved exactly and yield
\begin{subequations}
\label{eqs:idealflow}
\bea
\label{eq:backgroundv}
v_0(x) &= \frac{f(x)}{\sqrt{2}} \sqrt{1 \mp \sqrt{1-\alpha^2 f(x)^2}}, \\
\T_0(x) &= \frac{\sqrt{2} \, v_\infty \T_\infty}{\alpha f(x)} \sqrt{1 \pm\sqrt{1-\alpha^2 f(x)^2}},
\eea
\end{subequations}
where we have defined~$\alpha \equiv 2v_\infty\sqrt{1-v_\infty^2} \leq 1$.  Note the choice of sign: the fluid equations are satisfied for either choice, but consistency with the conditions at infinity requires that the upper (lower) signs be chosen when~$v_\infty < c_s$ ($v_\infty > c_s$), where~$c_s = 1/\sqrt{2}$ is the speed of sound\footnote{\label{foot:shocks}In fact, the fluid equations also allow ``shock wave" solutions where the choice of sign above changes abruptly, leading to a discontinuity in~$v_0(x)$ and~$\T_0(x)$ along the flow.}.  If $v_\infty = c_s$, both choices of sign give the correct conditions at infinity.

Requiring~$v_0$ and~$\T_0$ to be real imposes a constraint between the asymptotic velocity~$v_\infty$ and the maximum value~$f_\mathrm{max}$ of~$f(x)$:
\be
1-\alpha^2 f_\mathrm{max}^2 \geq 0 \Rightarrow \left(v_\infty^2 - c_s^2\right)^2 \geq \frac{f_\mathrm{max}^2 - 1}{4f_\mathrm{max}^2}.
\ee
Thus if~$f_\mathrm{max} > 1$, there exists a window of upstream velocities~$v_\infty$ centered around~$c_s$ for which~\eqref{eqs:idealflow} are ill-defined\footnote{In the gravitational context, this feature may be indicative of a regime in which a flowing black hole does not admit a hydrodynamic description.}.  To ensure that real solutions exist, we will therefore restrict our analysis to $f(x) \leq 1$.  That is, we take the fluid to flow over a gravitational potential ``well".

\subsection{Perturbation Equations}
\label{subsec:linpert}

Let us now consider linear perturbations of the ideal stationary flow~\eqref{eqs:idealflow}. Due to the translational invariance and stationarity of the background flow, we will decompose linear perturbations into Fourier modes~$e^{-i\omega t + i k y}$.  We thus write the flow as
\begin{subequations}
\label{eqs:perturbs}
\bea
u^\mu(t,x,y) &= u^\mu_0(x) + e^{-i\omega t + i k y}\left( \frac{v_0(x)}{f(x)^2} \, \delta v_x(x), \delta v_x(x), \delta v_y(x) \right), \\
\T(t,x,y) &= \T_0(x) \left(1 + e^{-i\omega t + i k y} \delta \T(x) \right), 
\eea
\end{subequations}
where $u^\mu_0$ and $\T_0$ are the stationary solutions~\eqref{eqs:idealflow}. Note that the perturbation of $u$ has been written so that it satisfies the linearized normalization condition $u^2 = -1$.

The linearized fluid equations are obtained by inserting~\eqref{eqs:perturbs} into the ideal fluid equations and expanding to linear order in the perturbation fields.  Using the equations of motion obeyed by the background flow~\eqref{eqs:idealflow}, we find that these linearized equations can be written in the form
\begin{subequations}
\label{eqs:idealpertEOM}
\bea
\delta v_x' - a_x(f) f' \, \delta v_x &= S_x, \\
\delta v_y' - a_y(f) f' \, \delta v_y &= S_y, \\
\delta \T' - a_\T(f) f' \, \delta v_x &= S_\T,
\eea
\end{subequations}
where the coefficients~$a_i(f)$ are algebraic functions of the well profile~$f(x)$, and the source terms~$S_x$,~$S_y$, and~$S_\T$ contain no derivatives and are linear in the perturbation fields.  These coefficients and source terms are given explicitly in Appendix~\ref{subapp:EOM}.

This system of equations~\eqref{eqs:idealpertEOM} can be rewritten in terms of a single third-order master equation for~$\delta v_x$:
\be
\label{eq:masterEOM}
\delta v_x'''(x) + q_1(x) \, \delta v_x''(x) + q_2(x) \, \delta v_x'(x) + q_3(x) \, \delta v_x(x) = 0,
\ee
where the coefficients~$q_i(x)$ contain~$k$,~$\omega$, and~$f(x)$ and its derivatives.  A solution to the above equation then allows us to reconstruct~$\delta v_y$ and~$\delta \T$ from~$\delta v_x$ and its first two derivatives.

For a given wave number~$k$,~\eqref{eqs:idealpertEOM} define an eigenvalue problem for the frequencies~$\omega$.  Because the equations~\eqref{eqs:idealpertEOM} are symmetric under either of the two transformations
\begin{subequations}
\bea
&k \to -k, \qquad \delta v_y \to -\delta v_y, \\
&\omega \to -\omega^*, \qquad \delta v_x \to \delta v_x^*, \qquad \delta v_y \to -\delta v_y^*, \qquad \delta \T \to \delta \T^*,
\eea
\end{subequations}
we may without loss of generality consider only~$k \geq 0$ and~$\Re(\omega) \geq 0$.  We now study this system for two potentials~$f(x)$: a toy model step well which yields analytically tractable results, and a smooth well which we approach numerically.

\subsection{The Step Well}
\label{subsec:squarewell}

Take the potential~$f(x)$ to be a piecewise constant step well:
\be
\label{eq:squarewell}
f(x) = \begin{cases} 1, & x < 0 \mbox{ or } x > L, \\ \fb, & 0 < x < L, \end{cases}
\ee
with~$\fb < 1$ so that the Newtonian potential~$\fb -1$ is negative.  This potential has infinite gradients in $f(x)$, so it isn't properly described by a gradient expansion. Nevertheless, it can be used as a simple toy model to understand the QNM spectrum of the equations~\eqref{eqs:idealpertEOM}.  Moreover, as mentioned in footnote~\ref{foot:shocks}, there are discontinuous fluid solutions which are well-described by ideal hydrodynamics; the model being studied here is qualitatively similar to these shock waves.  We will focus first on regions of constant~$f(x)$, then on how to stitch the solutions across the discontinuities in~$f(x)$.
\\\\
\noindent\underline{\textit{Constant~$f(x)$}}\\

\noindent Here, we concentrate on the middle region where $f(x) = \fb$; set~$\fb = 1$ for the other regions.  We will dress any objects associated with regions of constant~$f(x) = \fb$ with an overbar; the same objects with no overbar will correspond to regions where~$f(x) = 1$.  

\begin{figure}[t]
\centering
\subfigure[]{
\includegraphics[width=0.25\textwidth,page=8]{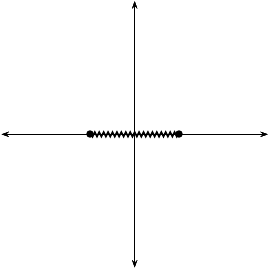}
\label{subfig:subsoniccuts}}
\hspace{2cm}
\subfigure[]{
\includegraphics[width=0.25\textwidth,page=9]{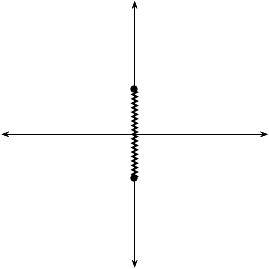}
\label{subfig:supersoniccuts}}
\caption{The branch cut structure of~$\Kb^\pm$ in the complex~$\omega$-plane.  \subref{subfig:subsoniccuts} and \subref{subfig:supersoniccuts} show the subsonic ($\vb < \cb_s$) and supersonic ($\vb > \cb_s$) cases, respectively.}
\label{fig:fluidcuts}
\end{figure}

Setting~$f(x) = \fb$, the master equation~\eqref{eq:masterEOM} reduces to
\begin{multline}
\label{eq:constantf}
\delta v_x'''(x) - \frac{(\fb^2-4\vb^2)i\omega}{\vb(\fb^2-2\vb^2)} \, \delta v_x''(x) + \frac{(4\fb^2-\vb^2)\omega^2 - \fb^2(\fb^2-\vb^2)k^2}{\fb^2(\fb^2-2\vb^2)} \, \delta v_x'(x) \\ + \frac{\left[\fb^2(\fb^2-\vb^2)k^2 - (2\fb^2-\vb^2)\omega^2\right]i\omega}{\fb^2 \vb (\fb^2-2\vb^2)} \, \delta v_x(x) = 0,
\end{multline}
where~$\vb$ is the constant background fluid velocity, which is related to the asymptotic fluid velocity via~\eqref{eq:backgroundv}.  Any solution to~\eqref{eq:constantf} can be written as
\be
\delta v_x(x) = A^+ e^{i \Kb^+ x} + A^- e^{i \Kb^- x} + A^0 e^{i \Kb^0 x},
\ee
where~$A^\pm$ and~$A^0$ are constants of integration, and we have also defined the wave numbers
\begin{subequations}
\label{eqs:dispersions}
\bea
\label{eq:pmdispersion}
\Kb^\pm &= \frac{(\pm \nub - \vb)\omega}{\fb^2 - 2\vb^2}, \mbox{ with } \nub = \sqrt{2} (\fb^2-\vb^2)\sqrt{\frac{1}{\fb^2}-\left(\frac{\cb_s^2-\vb^2}{\fb^2-\vb^2}\right) \frac{k^2}{\omega^2}}, \\
\label{eq:zerodispersion}
\Kb^0 &= \frac{\omega}{\vb},
\eea
\end{subequations}
where the speed of sound in this region is~$\cb_s \equiv \fb/\sqrt{2}$.  From our discussion in Section~\ref{sec:QNM}, the branch cut structure of~$\Kb^\pm$ in the complex~$\omega$-plane will determine the analytic structure of the retarded Green's function.  We therefore highlight the fact that for~$k \neq 0$,~$\nub$ (and thus~$\Kb^\pm$) has branch points in the complex~$\omega$-plane at
\be
\label{eq:nubranch}
\bar{\omega}_\mathrm{branch}(k) = \pm \fb \sqrt{\frac{\cb_s^2-\vb^2}{\fb^2-\vb^2}} \, k.
\ee
These branch points are analogous to those at~$\omega = \pm m$ for the massive scalar field discussed in Section~\ref{sec:QNM}.  For~$\vb < \cb_s$, the~$\omega_\mathrm{branch}(k)$ are real, and as for the scalar field we take the corresponding branch cut to connect them along the real axis.  For~$\vb > \cb_s$, the~$\omega_\mathrm{branch}(k)$ are imaginary, and we take the branch cut to connect them along the imaginary axis.  We illustrate these cuts in Figure~\ref{fig:fluidcuts}.

The wave numbers~\eqref{eqs:dispersions} define the allowed dispersion relations.  Indeed, for~$\vb = 0$, we find
\be
\label{eq:idealdisp}
\omega^2 = \cb_s^2 \, \vec{k}^2, \mbox{ with } \vec{k}^2 \equiv \left(\Kb^\pm\right)^2 + k^2,
\ee
which is just the usual dispersion relation for perturbations of the homogeneous ideal fluid on flat space, and indicates the existence of two sound modes with phase velocity equal to the speed of sound:
\be
v_\mathrm{phase}^\pm = \frac{\omega}{|\vec{k}|} = \pm \cb_s.
\ee
Similarly, the divergence of~$K^0$ at~$\vb = 0$ is indicative of a zero mode~$\omega = 0$ with phase velocity~$v_\mathrm{phase}^0 = 0$, as can be checked explicitly from the equations~\eqref{eqs:idealpertEOM}\footnote{Note that this is also the reason that~$\Kb^\pm$ become singular when~$\vb = \mp \cb_s$: the sound modes (which have phase velocities~$v_\mathrm{phase}^\pm = \pm \cb_s$ at~$\vb = 0$) become zero modes when boosted to a frame where the background fluid velocity is~$\vb = \mp \cb_s$.}.  The dispersion relations~\eqref{eqs:dispersions} at~$\vb \neq 0$ can then be obtained from these three modes via a Lorentz boost.  For convenience, we will continue to refer to~$e^{i K^0 x}$ as a ``zero mode'' even at nonzero velocity.
\\\\
\noindent\underline{\textit{Stitched Solutions}}\\

\noindent Next, let us stitch our solutions together.  At a discontinuity of~$f(x)$,~$f'(x)$ behaves as a delta function, and thus by~\eqref{eqs:idealpertEOM} the perturbations will be discontinuous as well.  We find that the fluid fields on the left~($-$) and right~($+$) sides of a discontinuity are related by the stitching conditions
\be
\label{eq:stitching}
\frac{\delta v_x^+}{\delta v_x^-} = \Delta_x, \qquad \frac{\delta v_y^+}{\delta v_y^-} = \Delta_y, \qquad \frac{\delta \T^+- \delta \T^-}{\delta v_x^-} = \Delta_\T,
\ee
where the~$\Delta_i$ depend only on~$v_\infty$ and the values~$f^\pm$ of~$f(x)$ on either side of the discontinuity.  We provide a derivation of these conditions and more explicit expressions in Appendix~\ref{subapp:stitching}.

We may now construct full solutions to the linearized equations in the step well~\eqref{eq:squarewell}.  These solutions will take the form
\be
\label{eq:squarewelldeltavx}
\delta v_x(x) = \begin{cases} A_L^+ e^{i K^+ x} + A_L^- e^{i K^- x} + A_L^0 e^{i K^0 x}, & x < 0 \\
A_M^+ e^{i \Kb^+ x} + A_M^- e^{i \Kb^- x} + A_M^0 e^{i \Kb^0 x}, & 0 < x < L \\
A_R^+ e^{i K^+ x} + A_R^- e^{i K^- x} + A_R^0 e^{i K^0 x}, & x > L
\end{cases},
\ee
where the subscripts~$L$,~$M$, and~$R$ stand for left, middle, and right of the well, respectively, and the unbarred wave numbers~$K^{\pm,0}$ can be found from~\eqref{eqs:dispersions} by setting~$\fb = 1$.  The stitching conditions~\eqref{eq:stitching} yield linear equations relating the~$A_i^{\pm,0}$ on either side of the discontinuities at~$x = 0$ and~$x = L$, and therefore allow us to construct a transfer matrix relating the coefficients on one side of the well to those on the other.  We find that
\be
\label{eq:transfer}
\begin{pmatrix} A_R^+ \\ A_R^- \\ A_R^0 \end{pmatrix} = 
\begin{pmatrix} M_{++} & M_{+-} & M_{+0} \\
M_{-+} & M_{--} & M_{-0} \\
0 & 0 & M_{00}  \end{pmatrix}
\begin{pmatrix} A_L^+ \\ A_L^- \\ A_L^0 \end{pmatrix},
\ee
where the~$M_{ij}$ depend on~$k$,~$\omega$,~$\fb$,~$v_\infty$, and~$L$.  Note that two entries of the transfer matrix vanish, implying that the coefficients~$A_i^0$ decouple from the other two.  As we will discuss later, this decoupling occurs for any~$f(x)$, even if $f(x)$ is not piecewise constant.

As an interesting aside, note that the exact solution~\eqref{eq:squarewelldeltavx} only has branch points in the complex~$\omega$-plane at the branch points of~$K^\pm$ and~$\Kb^\pm$. In particular, unlike the massive scalar field reviewed in Section~\ref{sec:QNM}, there is no branch point at~$\omega = 0$.  Thus the retarded Green's function for the square well, which inherits the branch cut structure of the solutions~\eqref{eq:squarewelldeltavx}, also does not exhibit a branch cut along the negative imaginary axis.  This is most likely due to the fact that~$f(x)$ differs from unity only in a region of compact support. Indeed, the scalar field Green's function~\eqref{eq:Greensfunction} only exhibits a branch cut along the imaginary axis whenever the potential~$V(x)$ does not vanish identically at sufficiently large~$x$~\cite{KokSch99}.  Thus this feature of our Green's function should be thought of as an artifact of the square well toy model being studied here, and is not expected to hold for more general~$f(x)$.
\\\\
\noindent\underline{\textit{Quasi-normal modes for subsonic flow}}\\

\begin{figure}[t]
\centering
\subfigure[]{
\includegraphics[page=10,width=0.47\textwidth]{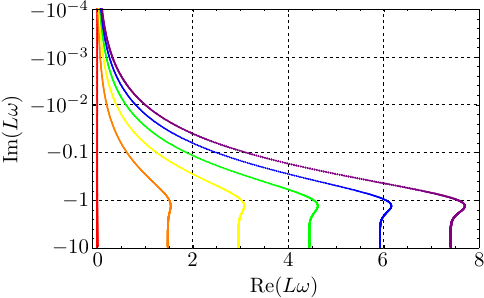}
\label{subfig:k0squarewellQNMsf}}
\hspace{0.4cm}
\subfigure[]{
\includegraphics[page=11,width=0.45\textwidth]{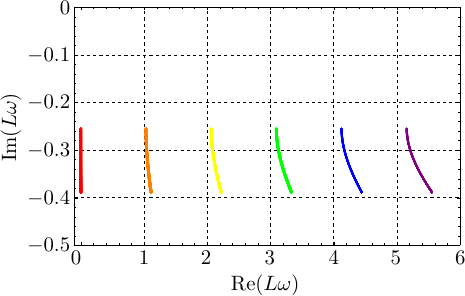}
\label{subfig:k0squarewellQNMsv}}
\caption{The first six QNMs~$\omega_n$ of the square well at vanishing transverse momentum~$k=0$ as the fluid velocity~$v_\infty$ and well depth~$\fb$ are varied.  From left to right, the curves represent~$n = 0,1,2,3,4,5$, respectively.  \subref{subfig:k0squarewellQNMsf}: the~$k = 0$ modes at fixed~$v_\infty = 0.5$ and varying~$\fb \in (0.0025,1-2\cdot 10^{-9})$.  Note the logarithmic scale for~$\Im(L\omega)$. As~$\fb$ is increased, the imaginary part~$\Im(L\omega)$ decreases; it becomes parametrically large and negative as~$\fb \to 1$ and small and negative as~$\fb \to 0$.  \subref{subfig:k0squarewellQNMsv}: the~$k = 0$ modes at fixed~$\fb = 0.5$ and varying~$v_\infty \in (0.001,0.7)$. As~$v_\infty$ is increased, the imaginary parts of the modes increases as well, but remains finite and nonzero as both~$v_\infty \to 0$ and as~$v_\infty \to 1/\sqrt{2} \approx 0.707$.}
\label{fig:k0squarewellQNMs}
\end{figure}

\noindent We are now prepared to study the QNM spectrum of the step well.  Consider first the subsonic case~$v_\infty < c_s$.  From the discussion in Section~\ref{sec:QNM}, determining the boundary conditions requires understanding the behavior of the functions~$e^{iK^\pm x}$,~$e^{iK^0 x}$ when~$\Im(\omega) > 0$.  From~\eqref{eqs:dispersions} with~$v_\infty < c_s$ and~$\Im(\omega) > 0$, we find
\be
e^{iK^+ x}, \, e^{iK^0 x} \to 0 \mbox{ as } x \to \infty, \quad e^{iK^- x} \to 0 \mbox{ as } x \to -\infty.
\ee
Thus a QNM must behave like~$e^{i K^- x}$ at~$x \to -\infty$ and like a superposition of~$e^{i K^+ x}$ and~$e^{i K^0 x}$ at~$x \to \infty$\footnote{At~$k = 0$, the~$K^\pm$ become proportional to~$\omega$.  In this special case, these boundary conditions reduce to the usual outgoing ones.}.  We therefore must have
\be
A_L^+ = A_L^0=0, \qquad A_R^- = 0,
\ee
so that~\eqref{eq:transfer} becomes
\be
\begin{pmatrix} A_R^+ \\ 0 \\ A_R^0 \end{pmatrix} = 
\begin{pmatrix} M_{++} & M_{+-} & M_{+0} \\
M_{-+} & M_{--} & M_{-0} \\
0 & 0 & M_{00}  \end{pmatrix}
\begin{pmatrix} 0 \\ A_L^- \\ 0 \end{pmatrix}.
\ee
The QNMs are then defined as functions of~$v_\infty$,~$\fb$,~$L$, and~$k$ by the transcendental equation
\be
M_{--} = 0.
\ee
This equation can be solved using any standard root-finding algorithm\footnote{For instance, \texttt{Mathematica}'s \texttt{FindRoot}.} to obtain the QNM spectrum.  Generically, we find that any~$\fb < 1$ yields an infinite family of modes at~$k = 0$ which take the form
\be
L\omega_n(k=0) = \beta n - i\alpha, \quad n \in \mathbb{Z},
\ee
where~$\alpha$ and~$\beta$ are positive constants that depend on~$v_\infty$ and~$\fb$.  In Figure~\ref{fig:k0squarewellQNMs}, we show the spectrum obtained at~$k = 0$ for various~$v_\infty$ and~$\fb$.  Specifically, in Figure~\ref{subfig:k0squarewellQNMsf} we fix~$v_\infty$ and vary~$\fb \in (0.0025,1-2\cdot 10^{-9})$, while in Figure~\ref{subfig:k0squarewellQNMsv} we fix~$\fb = 0.5$ and vary~$v_\infty \in (0.001,0.7)$.

Note the unusual feature that the imaginary part of the modes is independent of the overtone number~$n$; typically,~$\Im(\omega_n)$ decreases (that is, becomes more negative) with increasing~$n$.  We suspect the cause of this behavior is the infinitely sharp transition in~$f(x)$ at the walls of the well.  In a sense, the characteristic length scale~$\ell$ over which~$f(x)$ is varying is zero, so all the modes have vanishing period relative to this length scale (\textit{i.e.}~$\Re(\omega_n \ell) = 0$).  Since one would expect the damping of a mode to be determined by its interaction with the background, this implies that all the modes should have the same imaginary part.  This explanation cannot be complete, however, as it does not take into account the nonzero width~$L$ of the well.

Next, notice that while the~$k = 0$ modes do not vary substantially as the background fluid velocity~$v_\infty$ is varied, varying the well depth~$\fb$ changes the modes dramatically: as~$\fb \to 1$ (\textit{i.e.}~as the well becomes shallow and vanishes), the imaginary part of the modes becomes arbitrarily negative.  This must occur, as when the well disappears at~$\fb = 1$, QNMs cease to exist.  Similarly, as the well approaches its maximum depth~$\fb \to 0$, the modes become parametrically small.  In particular, they become arbitrarily long-lived, a phenomenon which we interpret as the effect of the well ``trapping'' modes inside it and preventing them from leaking out to infinity.

\begin{figure}[t]
\centering
\subfigure[]{
\includegraphics[page=12,width=0.45\textwidth]{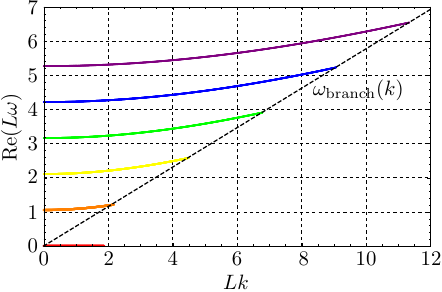}
\label{subfig:squarewellQNMsreal}}
\hspace{0.1cm}
\subfigure[]{
\includegraphics[page=13,width=0.485\textwidth]{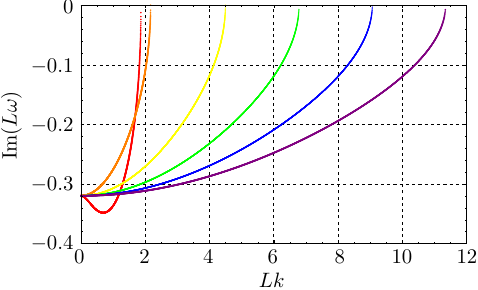}
\label{subfig:squarewellQNMsimag}}
\caption{The real and imaginary parts of the first six QNMs~$\omega_n$ of the square well for~$v_\infty = 0.5$,~$\fb = 0.5$ and varying~$k$.  From red to purple (bottom to top in the left figure, left to right in the right figure), the lines correspond to~$n = 0,1,2,3,4,5$.  The dashed line in the left figure denotes the location of the branch point~$\omega_\mathrm{branch}(k)$ of the asymptotic dispersion relations~$K^\pm(\omega,k)$.  Note that all the modes~$n \neq 0$ hit this branch point, while the pure imaginary~$n = 0$ mode runs into the branch cut along the real axis.}
\label{fig:squarewellQNMs}
\end{figure}

Finally, note the interesting feature that the~$n = 0$ mode is purely damped.  Generically, purely damped modes would lie on a branch cut along the negative imaginary axis, and therefore would not contribute as a QNM to the inverse Laplace transform~\eqref{eq:inverselaplace}.  But since the Green's function for the square well does not exhibit this branch cut, we allow purely damped QNMs.  This feature should be thought of as an artifact of the piecewise constant potential studied here, and is not expected to be reproduced by a generic potential.

In Figure~\ref{fig:squarewellQNMs}, we show the effect of increasing~$k$ for the choice~$v_\infty = 0.5$,~$\fb = 0.5$ (other parameters behave similarly).  Note that the pure damped mode ($n = 0$, red in Figure~\ref{fig:squarewellQNMs}) runs into the real axis, at which point it hits the branch cut shown in~Figure~\ref{subfig:subsoniccuts}.  Similarly, the~$n \neq 0$ modes also drift towards the real axis, but they instead reach the real axis precisely at the branch points~$\omega_\mathrm{branch}(k)$ of the asymptotic dispersion relations~$K^\pm(\omega,k)$.  Once a mode hits a branch cut or branch point, we no longer follow it onto a secondary Riemann surface.

Thus we find that for any background (\textit{i.e.}~for any choice of~$v_\infty$ and~$\fb$), there are modes whose imaginary parts are arbitrarily small, and are therefore arbitrarily long-lived.  Furthermore, these modes contact the real line at a finite wavenumber $k$. We will comment further on the importance of these modes in Section~\ref{sec:discussion}.
\\\\
\noindent\underline{\textit{Quasi-normal modes for supersonic flow}}\\

\noindent Now consider the supersonic case~$v_\infty > c_s$, where for~$\Im(\omega) > 0$ we find
\be
e^{iK^+ x}, \, e^{iK^0x}, \, e^{i K^- x} \to 0 \mbox{ as } x \to \infty.
\ee
From~\eqref{eq:generalGreens}, this implies that the retarded Green's function~$G(x;x')$ vanishes when~$x < x'$.  This is precisely what we would expect on physical grounds: the Green's function gives the response to a delta-function source at~$x = x'$, but if the flow is supersonic, any fluid excitations will be ``swept downstream'' with the fluid, and thus the region~$x < x'$ upstream of the perturbation cannot be affected.  The upshot of this observation is that the Wronskian~$W(x')$ can never vanish, and therefore there are no QNMs at all.

\subsection{A Smooth Well}
\label{subsec:smoothwell}

The step well has infinite gradients, and therefore is not expected to be well-described by a gradient expansion. We therefore now consider a smooth well, which allows us to keep gradients small.  Specifically, we take a Lorentzian:
\be
\label{eq:smoothf}
f(x) = 1 - \frac{a}{1+x^2/L^2}.
\ee
In the above, we can interpret~$\Phi_N(x) = -a/(1+x^2/L^2)$ as the Newtonian potential. As mentioned in Section~\ref{subsec:stationary}, we will take~$a > 0$ so that a background flow exists for all~$v_\infty$.  We consider only subsonic flows, since supersonic flows admit no QNMs (as discussed at the end of Section~\ref{subsec:squarewell}).

The setup here is similar to the step well.  At the asymptotic boundaries~$x \to \pm \infty$,~$f(x) \to 1$ so the solutions to the master equation~\eqref{eq:masterEOM} become
\be
\delta v_x(x\to \pm \infty) = A_{\pm\infty}^+ e^{i K^+ x} + A_{\pm\infty}^- e^{i K^- x} + A_{\pm\infty}^0 e^{i K^0 x},
\ee
where recall that the wave numbers~$K^{\pm,0}$ are given by~\eqref{eqs:dispersions} with~$\fb = 1$.  As discussed in Section~\ref{subsec:squarewell}, QNM boundary conditions require
\be
\label{eq:idealBC}
A^+_{-\infty} = A^0_{-\infty}=0, \qquad A^-_{+\infty} = 0.
\ee
Recall that in the step well, the mode proportional to~$A^0$ decouples from the other two, so that requiring~$A^0$ to vanish on one side of the well forces it to vanish on the other as well.  In Appendix~\ref{subapp:eigfuncs}, we show that this decoupling occurs for arbitrary~$f(x)$ by rewriting the fluid equations in a basis of asymptotic eigenfunctions. Thus the condition~$A^0_{-\infty} = 0$ implies~$A^0_{+\infty} = 0$ as well, and the boundary conditions become
\be
\label{eq:idealBCpsi}
\delta \Psi_i(x \to \pm\infty) \propto e^{i K^\pm x},
\ee
where for notational convenience we have introduced the perturbation vector~$\delta \Psi_i(x) \equiv (\delta v_x(x), \delta v_y(x), \delta \T(x))$.  The QNMs are thus defined as those~$\omega$ which admit solutions to the fluid equations~\eqref{eqs:idealpertEOM} which satisfy the boundary conditions~\eqref{eq:idealBCpsi}.

We solve for these modes numerically. We begin by changing to a new coordinate~$\xi$ defined via
\be
\label{eq:xidef}
\frac{x}{L} = \frac{\xi}{1-\xi^2},
\ee
so that $\xi\in(-1,1)$.  To impose the boundary conditions~\eqref{eq:idealBCpsi} we redefine the fluid variables via\footnote{Note that some redefinition of this kind is necessary since it is infeasible to control an exponential falloff numerically.}
\be
\label{eq:psidef}
\delta \Psi_i(\xi) = \exp\left[\frac{i}{2}\bigg((1-\xi)LK^-(\omega) + (1+\xi)LK^+(\omega) \bigg)\frac{\xi}{1-\xi^2}\right] \delta \psi_i(\xi).
\ee
With these redefinitions, the boundary conditions require that~$\delta \psi_i$ be constant on either boundary.  In particular, we can use the explicit asymptotic expressions~\eqref{eqs:fullsols} to constrain the~$\delta \psi_i$ at each boundary up to a single normalization constant. Specifically, we require
\be
\delta \psi_i(\pm 1) \propto \begin{pmatrix} k^2-2(1-v_\infty^2)\omega^2 \\
-\omega k (2v_\infty(1-v_\infty^2) \pm \nu) \\
-\sqrt{1-v_\infty^2} \, (k^2 v_\infty^2 \pm \omega^2 \nu) \end{pmatrix}.
\ee

While the original fluid equations in~$\delta \Psi_i(\xi)$ are linear in~$\omega$, the redefinition~\eqref{eq:psidef} yields equations for~$\delta \psi_i(\xi)$ that are non-polynomial in~$\omega$.  This prevents us from using standard linear eigenvalue solvers to obtain the QNM spectrum. We are fortunately aided by two special cases where the nonlinear eigenvalue problem becomes linear: (i) when~$k = 0$ (so that~$\nu = \sqrt{2}\, (1-v_\infty^2)$), and (ii) when we enforce that the modes coincide with branch points of~$\nu$, so that~$\nu = 0$.  

\begin{figure}[t]
\centering
\subfigure[]{
\includegraphics[page=14,width=0.45\textwidth]{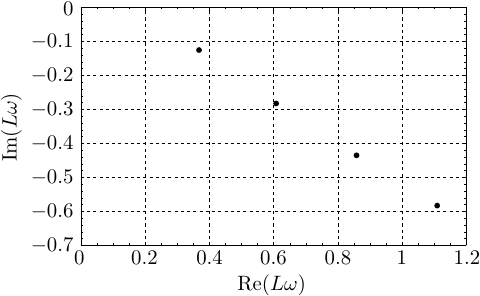}
\label{subfig:smoothwellQNMsk0}}
\hspace{0.5cm}
\subfigure[]{
\includegraphics[page=15,width=0.455\textwidth]{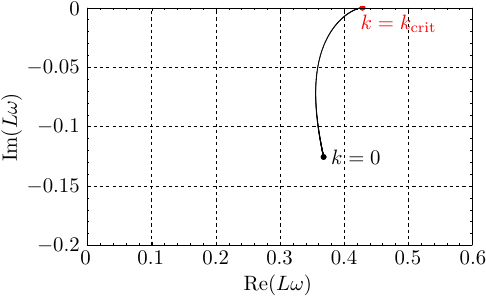}
\label{subfig:smoothwellQNMs}}
\caption{The first few quasinormal modes of the ideal fluid with~$f(x)$ as given in~\eqref{eq:smoothf}. Here, we take~$v_\infty = 0.55$ and~$a = 0.85$.  \subref{subfig:smoothwellQNMsk0}: the four lowest QNMs at~$k = 0$, computed by solving the linear eigenvalue problem (higher modes continue along the diagonal).  \subref{subfig:smoothwellQNMs}: the lowest QNM for nonzero~$k$, computed using Newton-Raphson. As for the square well, the mode tends towards the real axis until~$Lk_\mathrm{crit} \approx 0.806$, when it hits the branch point~$L\omega_\mathrm{branch}(k_\mathrm{crit}) \approx 0.429$.  This critical mode can also be obtained independently by solving a linear eigenvalue problem; the result is shown in red.  The higher modes behave similarly, though they may ``spiral'' around before approaching the real axis.}
\label{fig:idealQNMs}
\end{figure}

Our approach is thus as follows:
\begin{enumerate}
	\item At~$k = 0$, the equations for~$\delta \psi_i(\xi)$ yield a linear eigenvalue problem for~$\omega$, which we solve using standard eigenvalue methods (pseudospectral collocation on a Chebyshev grid and QZ factorization).  This yields a QNM spectrum~$\omega_n(k=0)$.
	\item For~$k \neq 0$ and~$\nu \neq 0$, we solve the nonlinear eigenvalue problem for the QNMs~$\omega_n(k)$ using Newton-Raphson (discretized in the same way as Step~1), taking the~$\omega_n(k=0)$ obtained in Step~1 as seeds.
	\item For modes that coincide with branch points of~$\nu$, we replace~$\omega$ with~$\omega_\mathrm{branch}(k)$ (\textit{\textit{c.f.}}~equation~\eqref{eq:nubranch}) everywhere in the equations for~$\delta \psi_i(\xi)$.  This sets~$\nu = 0$ and yields a linear eigenvalue problem for~$k$, we which we solve using the same methods of Step~1 to obtain a set of~$k_n$ corresponding to real QNMs~$\omega_n(k_n) = \omega_\mathrm{branch}(k_n)$.
\end{enumerate}
For more details on these numerical algorithms, see~\textit{e.g.}~\cite{Dias:2015nua}.

In Figure~\ref{fig:idealQNMs}, we show the QNM spectrum we obtained for the parameters~$v_\infty = 0.55$,~$a = 0.85$, for which the numerics converged particularly well (other parameter choices yield qualitatively similar results).  At~$k = 0$, all the modes are stable, like in the step well.  However, unlike the step well, the damping~$\Im(\omega_n)$ decreases for increasing~$n$.  As~$k$ is increased, this damping decreases until at~$Lk = Lk_\mathrm{crit} \approx 0.806$ the lowest mode hits the branch point~$L\omega_\mathrm{branch}(k_\mathrm{crit}) \approx 0.429$.  This critical mode can be obtained independently via Step~3 outlined above.  The higher modes behave similarly, though tracking them to the real axis requires using higher numerical precision.


\section{Viscous Effects}
\label{sec:viscousfluid}

We have shown that the ideal fluid equations linearized around the background~\eqref{eq:idealansatz} exhibit a continuum of QNMs which are arbitrarily long-lived.  Now, recall that our motivation was to study the potential fluid-like instabilities of flowing black holes.  However, not all stationary ideal fluid flows can be mapped to stationary black hole solutions. Essentially, this is because ideal fluids do not generate entropy, while flowing black holes necessarily do\footnote{To leading nontrivial order, the divergence of the fluid entropy current~$J^a_s$ is given by
\be
\grad_a J^a_s \sim \sigma_{ab} \sigma^{ab},
\ee
where~$\sigma_{ab}$ is the shear of the fluid flow.  This divergence is second-order in gradients, and thus is not captured by the ideal fluid equations (which are only first-order in gradients).}.  For example, on flat space, the stationary laminar flow
\be
u_0^\mu = \frac{1}{\sqrt{1-v_0(y)^2}} \left(1, v_0(y), 0\right)
\ee
is a solution to the ideal fluid equations for any velocity profile~$v_0(y)$.  Viscous corrections, however, will require~$v_0(y)$ to be constant.  Physically, this is a result of the fact that nontrivial~$v_0(y)$ generates shear and thus entropy at higher orders.  Since any gravitational solution will be dissipative, its ideal fluid dual must therefore arise as the zero-viscosity (or high-temperature) limit of a viscous solution.

\begin{figure}[t]
\centering
\subfigure[]{
\includegraphics[page=16,width=0.45\textwidth]{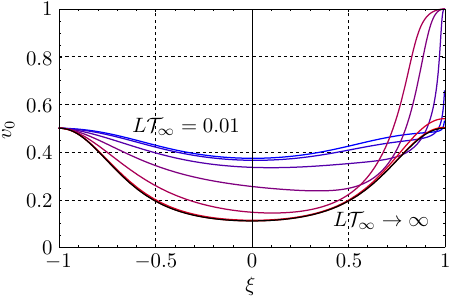}
\label{subfig:firstorderfluidv}}
\hspace{0.5cm}
\subfigure[]{
\includegraphics[page=17,width=0.45\textwidth]{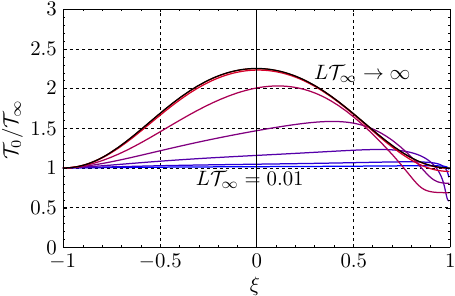}
\label{subfig:firstorderfluidT}}
\caption{Solutions to the first-order fluid equations for the background flow~\eqref{eq:idealansatz} for a fixed incoming fluid velocity~$v_\infty = 0.5$ and well depth~$a = 0.5$. Note that we use the compactified spatial coordinate~$\xi$ (defined in~\eqref{eq:xidef}).  Colors from blue to red (top to bottom in left, bottom to top in right) indicate increasing incoming fluid temperature. Here we show~$L\T_\infty = 0.01,0.03,0.1,0.3,1,10,100$.  Note that as~$L\T_\infty \to \infty$, the flow approaches the ideal solution~\eqref{eqs:idealflow}, shown in black.}
\label{fig:viscousbackground}
\end{figure}

Thus in order to ensure that the fluid analysis presented in Section~\ref{sec:idealfluid} captures the high-temperature limit of a dual flowing black hole, we must confirm that the ideal background flow~\eqref{eq:idealansatz} can be thought of as the inviscid limit of a viscous flow.  To that end, in Figure~\ref{fig:viscousbackground} we show numerical solutions to the first-order fluid equations with~$f(x)$ as given in~\eqref{eq:smoothf} at constant~$v_\infty = 0.5$,~$a = 0.5$ and different values of~$L\T_\infty$ from~$0.01$ to~$100$.  These first-order equations are obtained by including the dissipative term
\be
\Pi^{ab} = -2\T^2 \sigma^{ab}
\ee
in~\eqref{eq:Tconstitutive}, with~$\sigma^{ab}$ the shear of the fluid flow (\textit{i.e.}~the traceless symmetric part of~$\grad^a u^b$).  Note that as~$L\T_\infty \to \infty$, the viscous solutions approach the ideal solution~\eqref{eqs:idealflow}, and thus our ideal background fluid should well-approximate the high-temperature limit of a flowing black hole.

However, the linear dynamics of viscous flows may exhibit novel behavior not present in the strict inviscid limit.  In particular, viscous terms may cause new QNMs to appear which are qualitatively different from the ideal modes found above~\cite{Cardoso:2009bv}.  Additionally, recall that the ideal fluid modes run into the branch points~$\omega_\mathrm{branch}$ at finite momentum~$k$.  Viscous corrections may change this behavior, so that the corrected modes may approach a vicinity of the branch points but never be absorbed by them.  The modes may then have a continuation for larger~$k$ whose behavior would not be continuously connected to any of the ideal modes.

It would therefore be interesting to generalize our analysis to the viscous fluid equations.  Unfortunately, the difficulty involved in a viscous analysis renders it outside the scope of this Paper.  Here we comment on some obstacles:
\begin{itemize}
	\item It is well-known that the first-order fluid equations do not yield a well-posed hyperbolic problem, and in fact can give rise to unphysical instabilities and acausality~\cite{Isr76,HisLin83,HisLin85,HisLin87} (this does not prevent us from constructing \textit{stationary} solutions like those in Figure~\ref{fig:viscousbackground}, however).  In order to obtain a well-posed system of equations, one must resort to the second-order fluid equations, as described in \textit{e.g.}~\cite{GreCar13}.  Thus a consistent linear stability analysis must in fact study the linearization of the \textit{second-order} fluid equations.
	\item These second-order fluid equations cannot be solved analytically for the background flow~\eqref{eq:idealansatz}.  Without an analytic background, it is not possible to obtain toy model analytic solutions to the linearized equations (as we have done for the square well presented in Section~\ref{subsec:squarewell}).
	\item A numerical approach to solving the linearized second-order fluid equations must contend with the fact that they admit five independent dispersion relations, rather than the three obtained for the ideal fluid.  Because these five asymptotic behaviors do not decouple in any obvious way, generically a superposition of them is permitted in each asymptotic region~$x \to \pm \infty$.  The simple two-sided ansatz~\eqref{eq:psidef} is insufficient to capture such a superposition, and thus we would need to work in a basis of asymptotic eigenfunctions such as the~$g_i(x)$ introduced in Appendix~\ref{subapp:eigfuncs}.
	\item The five viscous dispersion relations are non-polynomial in~$\omega$, even when~$k = 0$.  It is therefore not possible to obtain a linear (or polynomial) eigenvalue problem for~$\omega$, and we would need to resort to Newton-Raphson to search for the QNMs.  Such a search would require finding good seed solutions, but the only readily available ones are the ideal fluid modes, which would not yield qualitatively new QNMs.
\end{itemize}

While we can say little about any new QNMs that appear due to viscous corrections, on physical grounds we should expect that viscosity should dampen the ideal modes shown in Figure~\ref{fig:idealQNMs}.  Indeed, one can check this by na\"ively proceeding as follows.  Of the five dispersion relations that appear in the linearized second-order fluid equations, two reduce to~$K^\pm$~\eqref{eq:pmdispersion} in the ideal fluid limit~$L\T_\infty \to \infty$.  By taking the ansatz~\eqref{eq:psidef} with~$K^\pm$ replaced by these second-order dispersion relations, we search for solutions to the linearized second-order fluid equations using Newton-Raphson with the ideal QNMs as seeds.  As expected, we find that at finite~$L\T_\infty$ the QNMs become more damped (\textit{i.e.}~their imaginary part decreases).  We emphasize, however, that this is far from a thorough treatment of the full second-order equations.  In particular, the other exponential behaviors are not under good numerical control.



\section{Discussion}
\label{sec:discussion}

In this Paper, we investigated the stability of flowing black holes by working in the hydrodynamic regime.  Our approach involved studying the linear stability of an ideal conformal fluid dual to the high-temperature limit of a flowing black hole.  Specifically, we considered laminar flows over gravitational potential wells.  For these flows, we found no unstable modes, but on any subsonic background flow we found a continuum of arbitrarily long-lived modes at finite transverse momentum~$k$ near~$k_\mathrm{crit}$ (and its analogs for higher overtone numbers~$n$).  Interestingly, these modes persist even in the limit where the background flow velocity vanishes, at least for the square well.

Our results are tantalizing: though we have found no \textit{linear} instability, these arbitrarily long-lived modes imply that nonlinear interactions may play an important role in a full dynamical evolution of the fluid equations.  In fact, as discussed in Section~\ref{sec:intro}, an Orr-Sommerfeld analysis of many solutions to the Navier-Stokes equation does not yield any unstable modes.  However, a fully nonlinear treatment (either through full time evolution of the Navier-Stokes equation or performing an experiment) finds that virtually all flows are unstable at sufficiently high~$Re$.  Since the ideal fluid has vanishing viscosity, it can formally be thought of as the limit~$Re \to \infty$, and thus would be expected to always be unstable.

In addition to nonlinear mechanisms, it is possible that linear but non-modal effects may play a role.  Indeed, a linear non-modal analysis of the nonrelativistic Navier-Stokes equation, wherein one studies the ``pseudospectrum'' of the linearized system, reveals that linear effects can lead to a large but finite amplification of small initial data.  We defer to~\cite{TreTre93} for details.  

Of course, the reader may note that the inclusion of viscosity might suppress any potential nonlinear interactions that would give rise to instabilities, since viscous corrections might render the lifetime of the ideal fluid modes finite.  But this would only be true at small~$Re$: since the ideal fluid emerges as the high-$Re$ (low-viscosity) limit of the viscous equations, we would expect the lifetime of these modes to be parametrically large in~$1/Re$.  Thus if nonlinear interactions do play an important role in the ideal fluid, they should still be present in the viscous equations at sufficiently high~$Re$.  This is entirely consistent with usual fluid dynamics, where flows are typically unstable only at sufficiently high~$Re$.  Similar arguments also apply to the viscous psuedospectrum: any damping would be expected to be parametrically small in~$1/Re$, and thus any non-modal amplification will still occur for sufficiently large~$Re$.

An analogy can be drawn to near-extremal Kerr black holes, which exhibit QNMs with characteristic timescales parametrically large in the inverse extremality parameter.  As argued (and demonstrated via a toy model) in~\cite{YanZim14}, a nonlinear parametric resonance instability between these long-lived QNMs of Kerr induces a turbulent-like cascade of energy between them.  It is conceivable that a similar phenomenon might occur in flowing black holes.

Of course, verifying whether nonlinear or non-modal effects play an important role requires an analysis that goes beyond that performed here.  In analogy with~\cite{YanZim14}, one might try to construct a toy model that captures the nonlinear inter-mode coupling to show how energy might be exchanged between long-lived modes.  However, a more proper treatment should also include viscous effects, which as discussed in Section~\ref{sec:viscousfluid} introduce substantial challenges.  For these reasons, we believe the most conclusive study would be a fully time-dependent numerical simulation of either the fluid flow or the gravitational solution itself.  This approach would have the added benefit of providing the end state of any potential instability, which could exhibit structure different from currently known gravitational instabilities.  We therefore leave this dynamical construction to future work.

We close with some remarks.  First, one may also be tempted to draw analogies to global AdS, where there are an infinite number of undamped modes which are generically suspected to lead to a nonlinear instability~\cite{DafermosHolzegel2006,Bizon:2011gg}.  However, this instability may be intimately tied to the fact that the spectrum is resonant.  Our system, unlike pure global AdS, is very different in that it involves high temperatures with large horizons, has a non-compact boundary metric, and likely does not have a resonant spectrum.

Second, while we focused only on computing QNMs, recall from the discussion in Section~\ref{sec:QNM} that branch cuts of the retarded Green's function also contribute to the time evolution of initial data.  Since the QNMs we have found eventually get ``absorbed'' by this branch cut, it would be interesting to study its contribution to the late-time behavior of fluid perturbations.

Finally, it is known~\cite{BhaMin08} that the non-relativistic, incompressible Navier-Stokes equations emerge as a scaling limit from the relativistic equations~\eqref{eq:Tconservation}.  One might therefore wonder if studying the non-relativistic scaling limit of our system might lead to further insights into its stability.  In fact, in this scaling limit, our system on the particular background~\eqref{eq:metric} reduces to a uniform, unforced fluid flow on flat space.  Such a system is clearly nonlinearly stable, and thus offers little new insight\footnote{The pessimistic reader might be concerned that such a trivial nonrelativistic limit might imply that the relativistic system from which it descends might be uninteresting as well.  But this is clearly not the case: our relativistic system exhibits a rich structure of QNMs, while a uniform fluid on flat space exhibits no QNMs at all.}.

\acknowledgments

It is a pleasure to thank Jyotirmoy Bhattacharya, Gavin Hartnett, Luis Lehner, Don Marolf, Mukund Rangamani, Chris Rosen, Jorge Santos, Andrew Walton, Toby Wiseman, and Alexander Zhidenko for useful discussions and correspondence.  SF is supported by the ERC Advanced grant No.~290456.  During the initial stages of this project, SF was also supported by the National Science Foundation under grant number PHY12-05500 and by funds from the University of California. B.W. is supported by European Research Council grant no. ERC-2011-StG 279363-HiDGR.

\appendix

\section{Fluid Perturbations}
\label{app:ideal}

In this Appendix, we present in more detail the linearized equations of motion for the ideal fluid discussed in the main text.

\subsection{Equations of Motion}
\label{subapp:EOM}

Using the equations of motion for the background velocity and temperature as well as the explicit expression~\eqref{eq:backgroundv} for the background velocity, the linearized ideal fluid equations can be written in the form
\begin{subequations}
\label{eqs:appEOM}
\bea
\label{eq:appdeltavx}
\delta v_x' - a_x(f) f' \, \delta v_x &= S_x, \\
\label{eq:appdeltavy}
\delta v_y' - a_y(f) f' \, \delta v_y &= S_y, \\
\label{eq:appdeltaT}
\delta \T' - a_\T(f) f' \, \delta v_x &= S_\T.
\eea
\end{subequations}
Explicitly, the~$a_i(f)$ and~$S_i$ are given by
\begin{subequations}
\label{eqs:as}
\bea
a_x(f) &= \frac{\alpha^2 f}{1-\alpha^2 f^2} + \frac{1}{f\sqrt{1-\alpha^2 f^2}}, \\
a_y(f) &= \frac{1}{2f}\left(1 + \frac{1}{\sqrt{1-\alpha^2 f^2}}\right), \\
a_\T(f) &= -\frac{\alpha(1 + \sqrt{1 - \alpha^2 f^2})}{2(1-\alpha^2 f^2)},
\eea
\end{subequations}
where recall that~$\alpha \equiv 2v_\infty\sqrt{1-v_\infty^2}$, and
\begin{subequations}
\begin{multline}
S_x(x) = \frac{1}{f(x)^2-2v_0(x)^2}\left[-i\omega v_0(x)\, \delta v_x(x) - i k f(x)^2\, \delta v_y(x) \phantom{2i\omega\sqrt{f(x)^2-v_0(x)^2}} \right. \\ \left. + 2i\omega\sqrt{f(x)^2-v_0(x)^2} \, \delta\T(x)\right],
\end{multline}
\be
S_y(x) = \frac{1}{v_0(x)}\left[i\omega \, \delta v_y(x) - ik\sqrt{f(x)^2-v_0(x)^2} \, \delta \T(x)\right],
\ee
\begin{multline}
S_\T(x) = \frac{1}{f(x)^2-2v_0(x)^2} \left[\frac{i\omega (f(x)^2-v_0(x)^2)^{3/2}}{f(x)^2} \, \delta v_x(x) \right. \\ \left. \phantom{\frac{i\omega (f(x)^2-v_0(x)^2)^{3/2}}{f(x)^2}} + ik v_0(x) \sqrt{f(x)^2 - v_0(x)^2} \, \delta v_y(x) - i\omega v_0(x) \, \delta \T(x) \right],
\end{multline}
\end{subequations}
where~$v_0(x)$ is the background local fluid velocity, given explicitly in terms of~$f(x)$ and~$v_\infty$ in~\eqref{eq:backgroundv}.

\subsection{Stitching Conditions}
\label{subapp:stitching}

A discontinuity of~$f(x)$ should be thought of as a limiting case of a smooth~$f(x)$.  In order to obtain stitching conditions on the fluid perturbations, we therefore consider integrating the equations~\eqref{eqs:idealpertEOM} across a sharp but smooth jump in~$f(x)$.  At such a jump (which for simplicity we will take to be at~$x = 0$),~$f'(x)$ will be large, and thus the terms on the right-hand side of equations~\eqref{eqs:appEOM} will not contribute.  For example, to obtain the discontinuity in~$\delta v_x(x)$, we divide~\eqref{eq:appdeltavx} by~$\delta v_x(x)$ and integrate across~$x = 0$:
\be
\int_{-\eps}^\eps \frac{\delta v_x'(x)}{\delta v_x(x)} \, \dd x - \int_{-\eps}^\eps a_x(f(x)) f'(x) \, \dd x = \int_{-\eps}^\eps \frac{S_x(x)}{\delta v_x(x)} \, \dd x.
\ee
Assuming that~$S_x(x)/\delta v_x(x)$ is finite near~$x = 0$, in the limit~$\eps \to 0$\footnote{The limit~$\eps \to 0$ is a double limit, as the jump in~$f(x)$ must become correspondingly sharper in order to remain within the region~$(-\eps, \eps)$.} the right-hand side of the above vanishes and we can integrate the terms on the left-hand side to obtain
\be
\frac{\delta v_x(0^+)}{\delta v_x(0^-)} = \exp\left[\int_{f_-}^{f_+} a_x(f) \, \dd f\right],
\ee
where~$f^\pm \equiv f(0^\pm)$ are the values of~$f(x)$ on the left and right sides of the discontinuity.  We obtain an analogous expression for the discontinuity in~$\delta v_y(x)$, while for~$\delta \T(x)$ we rearrange~\eqref{eq:appdeltaT} into
\be
\left(\frac{\delta \T(x)}{\delta v_x(x)}\right)' + \frac{\delta \T(x)}{\delta v_x(x)} \frac{\delta v_x'(x)}{\delta v_x(x)} - a_\T(f(x)) f'(x) = \frac{S_\T}{\delta v_x(x)},
\ee
after which we can use~\eqref{eq:appdeltavx} to replace~$\delta v_x'(x)/\delta v_x(x)$ and integrate as before to obtain
\be
\frac{\delta \T(0^+) - \delta \T(0^-)}{\delta v_x(0^-)} = \int_{f_-}^{f_+} a_\T(f) \exp\left[\int_{f_-}^f a_x(\tilde{f}) \, \dd \tilde{f}\right] \, \dd f.
\ee
Using the explicit expressions~\eqref{eqs:as}, our final stitching conditions thus become
\begin{subequations}
\bea
\frac{\delta v_x(0^+)}{\delta v_x(0^-)} &= \Delta_x \equiv \frac{f_+^5(f_-^2-2v_-^2)(f_-^2-v_-^2)}{f_-^5(f_+^2-2v_+^2)(f_+^2-v_+^2)}, \\
\frac{\delta v_y(0^+)}{\delta v_y(0^-)} &= \Delta_y \equiv \frac{f_+^2}{f_-^2} \sqrt{\frac{f_-^2-v_-^2}{f_+^2-v_+^2}}, \\
\frac{\delta \T(0^+) - \delta \T(0^-)}{\delta v_x(0^-)} &= \Delta_\T \equiv \frac{(f_+^2 v_-^2 - f_-^2 v_+^2)\sqrt{f_-^2-v_-^2}}{f_-^2 v_-(f_+^2 - 2v_+^2)},
\eea
\end{subequations}
where we have written the expressions in terms of the background velocities~$v_\pm = v_0(0^\pm)$, which via~\eqref{eq:backgroundv} are related to the asymptotic fluid velocity~$v_\infty$:
\be
v_\pm = \frac{f_\pm}{\sqrt{2}}\sqrt{1-\sqrt{1-4v_\infty^2(1-v_\infty^2)f_\pm^2}}.
\ee
Note that we have assumed~$v_\infty < c_s$ in choosing the sign inside the first square root.

\subsection{Decoupling of the Zero Mode}
\label{subapp:eigfuncs}

In order to show that the zero mode~$e^{i K^0 x}$ decouples from the other two, we re-express the fluid equations in terms of the asymptotic eigenfunctions corresponding to the three modes~$K^{\pm,0}$.  To do so, first note that in a region of constant~$f(x) = \fb$ (and therefore constant~$v_0(x) = \vb$), full solutions to~\eqref{eqs:idealpertEOM} can be written as
\begin{subequations}
\label{eqs:fullsols}
\be
\delta v_x(x) = \fb^2\Big(\fb^4 k^2 -2(\fb^2-\vb^2)\omega^2\Big)\left[A^+ e^{i \Kb^+ x} + A^- e^{i \Kb^- x} + A^0 e^{i \Kb^0 x}\right],
\ee
\begin{multline}
\delta v_y(x) = -\fb^2 \omega k\left[A^+ e^{i \Kb^+ x} \Big(2\vb(\fb^2-\vb^2) + \fb^2 \nub\Big) + A^- e^{i \Kb^- x} \Big(2\vb(\fb^2-\vb^2) - \fb^2 \nub\Big)\right] \\ + A^0 e^{i \Kb^0 x} \frac{\omega}{k \vb} (\fb^2-\vb^2)\Big(2(\fb^2-\vb^2)\omega^2 - \fb^4 k^2\Big),
\end{multline}
\be
\delta \T(x) = -\fb^2\sqrt{\fb^2-\vb^2} \left[A^+ e^{i \Kb^+ x}(\fb^2 k^2 \vb + \omega^2 \nub) + A^- e^{i \Kb^- x}(\fb^2 k^2 \vb - \omega^2 \nub)\right],
\ee
\end{subequations}
with~$\Kb^{\pm,0}$ and~$\nub$ defined as in~\eqref{eq:pmdispersion}.  For general~$f(x)$, we use the above expressions to define new fluid variables~$g_+(x)$,~$g_-(x)$, and~$g_0(x)$ by making the substitutions~$\fb \to f(x)$,~$\vb \to v_0(x)$, and~$A^{\pm,0} \to g_{\pm,0}(x)$.  Note that we also replace~$\vb$ and~$\fb$ in the expressions for~$\Kb^{\pm,0}$ and~$\nub$, so that
\begin{multline}
\Kb^\pm \to K^\pm(x) = \frac{\pm \nu(x) - v_0(x)}{f(x)^2-2v_0(x)^2}, \qquad \Kb^0 \to K^0(x) = \frac{\omega}{v_0(x)}, \\ \mbox{ with } \nu(x) = \left(f(x)^2-v_0(x)^2\right)\sqrt{\frac{2}{f(x)^2}-\left(\frac{f(x)^2-2v_0(x)^2}{f(x)^2-v_0(x)^2}\right)\frac{k^2}{\omega^2}}.
\end{multline}
We then re-express the fluid equations in terms of~$g_{\pm,0}(x)$.  Though this redefinition is nontrivial and consequently the resulting equations contain more terms, they are engineered to remove the source terms in~\eqref{eqs:idealpertEOM}:
\be
\label{eq:eigfuncs}
g_i' - M_{ij}(f) f' \, g_j = 0,
\ee
as when~$f'(x) = 0$ the~$g_i(x)$ must be constants by construction.

Note that~\eqref{eq:eigfuncs} should be thought of as a continuum version of~\eqref{eq:transfer}.  Indeed, while we will not provide the terms in~$M_{ij}(x)$ explicitly, we note that just as in the step well,~$M_{0+} = 0 = M_{0-}$.  Thus as advertised in the main text, the equation for~$g_0(x)$ decouples from the other two:
\be
g_0' - M_{00}(f) f' \, g_0 = 0.
\ee
In particular, the QNM boundary conditions~\eqref{eq:idealBC} imply that
\begin{subequations}
\bea
A_{-\infty}^+ = g_+(x \to -\infty) &= 0, \\
A_{-\infty}^0 = g_0(x \to -\infty) &= 0, \\
A_{+\infty}^- = g_-(x \to +\infty) &= 0.
\eea
\end{subequations}
Thus due to the decoupling of~$g_0(x)$, the second of these conditions implies that~$g_0(x) = 0$ everywhere, so that~$A_{+\infty}^0 = g_0(x \to +\infty) = 0$ as well.

\bibliographystyle{jhep}
\bibliography{biblio}

\end{document}